# Electromagnetic Waves Through Disordered Systems: Comparison Of Intensity, Transmission And Conductance

Gabriel Cwilich and Fredy R Zypman

Yeshiva University, Department of Physics, New York, NY 10033-3201

## ABSTRACT

We obtain the statistics of the intensity, transmission and conductance for scalar electromagnetic waves propagating through a disordered collection of scatterers. Our results show that the probability distribution for these quantities, x, follow a universal form $Y_U(x) = x^\alpha e^{-x^\mu}$. This family of functions includes the Rayleigh distribution (when $\alpha=0$, $\mu=1$) and the Dirac delta function ($\alpha \rightarrow +\infty$), which are the expressions for intensity and transmission in the diffusive regime neglecting correlations. Finally, we find simple analytical expressions for the $n^{th}$ moment of the distributions and for to the ratio of the moments of the intensity and transmission, which generalizes the n! result valid in the above regime.

## INTRODUCTION

Interest in wave propagation was rekindled by the work of Anderson[1] showing that in the presence of disorder the nature of the transport process can change radically from what was predicted by diffusion theory, to a localized regime in which the signal decays exponentially inside the medium. The problem of understanding and characterizing the different regimes has become, thus, central to the field[2], and has stimulated research in many different types of systems in a wide area of physics ranging from electronic materials[3,4], acoustic[5], optic and laser systems[6,7,8,9], microwaves[10] to atomic physics[11].

Since there is no complete agreement about the interpretation of experimental results[12] on the propagation in different regimes, the study of the statistical properties[13,14] of the transmittance quantities becomes of great interest, because they can be now measured with precision[15,16], and they convey information on the nature of the transport process inside the system[17]. When the wave travels through a disordered system the value of the intensity at a given location, depends on the exact position of the scatterers, which is only known in a statistical sense. It is common practice, then, to obtain the probability distribution of the intensity for an ensemble of configurations, and try to determine if the statistical properties of the distribution of the transmittance quantities (intensity, transmission and conductance) can be used as signatures of the various propagating regimes.

With this in mind we introduce here a simple model consisting of scalar waves incident on a system of randomly distributed scatterers, which act as sources of secondary radiation. We determine the distributions of the propagating intensity (one incident mode - one collected mode), the transmission (one incident mode- all outgoing modes) and the conductance (all incident modes- all outgoing modes) by numerically evaluating these quantities for different members of the ensemble of the disordered system, and we compare the moments of their distributions with theoretical predictions. To calculate the intensity of one particular sample, we set an incoming plane wave from a fixed direction. The collection of particles then scatters radiation from this incoming wave, and the outgoing wave is finally calculated at a fixed point in space. Then we generate another sample, belonging to the same ensemble and we recalculate the intensity. We repeat the process a large number of times (64,000) and construct the probability distribution of the intensity. We found that the probability distribution is extremely well estimated by a universal function[18] $Y_U(x) \propto x^\alpha e^{-x^\mu}$ where $x \equiv \frac{I}{\langle I \rangle}$. Similar studies were performed for Transmission and Conductance. In the case of Conductance we also considered the incoming wave from different directions to have random phase, to mimic the realistic situation of different incoherent waves incident on the sample. In all cases the probability distributions are well approximated by $Y_U(x)$.

**BASIC CONFIGURATION**

The system is randomized, and it is characterized by the filling fraction of volume of scatterers to total volume. A typical configuration is shown in Figure 1. An incoming plane wave with direction $\vec{k}_a$ impinges on the system. A generic detector is placed along the direction $\vec{k}_b$, where the scattered wave is measured. The square of the magnitude of the electric field for a fixed direction is the intensity $I_{ab}$, ($I_{ab}=|E_{ab}|^2$). The transmission is defined by $T_a=|\Sigma_b E_{ab}|^2$, that is, the intensity collected from many detectors. Finally, the conductance is $\sigma = |\Sigma_{ab} E_{ab}|^2$, corresponding to the experimental situation of many sources and many detectors.

We use Scalar Diffraction Theory, in which the field is given by Kirkchoff's expression $E = \sum \frac{e^{ikr}}{r}(\cos\beta + \cos\theta)$, $\beta$ being the angle of incidence of the plane wave, and $\theta$ the angle of emergence of the deflected wave, $k$ is the wave number, $r$ is the distance from a point on the surface of a scatterer to a given detector, and the sum runs over the surface of all the scatterers. We obtain numerically, for different filling fractions, the probability distribution of all the transmittance quantities.

This study is a follow up of a previous work[19]. In this case we have repeated the previous numerical study using an algorithm of generation of the random configurations that excludes the possibility of overlap among scatterers. Using the same size of the statistical ensemble as before[19], this study shows much smoother histograms of the transmittance quantities. This extreme smoothness allows us to improve the validation of any proposed analytical parametrization of the curves. Another new feature of this study is that, in the case of conductance, we have also randomized the phases of the incident waves from different directions, to make better contact with the quantities measured in electronic systems[20].

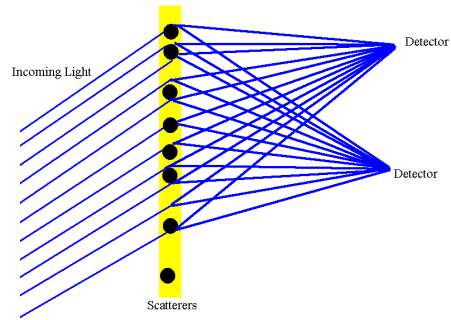

Figure 1. Scattering system.

All distributions are in very good agreement with $Y_U(x)$; in particular for all the cases of Intensity, Transmission and Conductance, and the lower concentrations of Random Phase Conductance the agreement is excellent. The best values for the parameters $\alpha$ and $\mu$ have been found for all cases, and some of them are reported in the next section.

**RESULTS**

We produced histograms of $I_{ab}$, $T_a$, and C, for filling factors between 10% and 90%. To build each histogram, we considered 64,000 different configurations of the ensemble. For each configuration, we calculated the desired transmittance quantity. Typical results for low and high concentration are shown in figure 2. There we show the numerical results (dots) and the best fit to $Y_U(x)$ (full line). The parameter $\mu$ is generally connected to the tail of the distribution (smaller $\mu$ corresponding to longer tails) while the parameter $\alpha$ is related to the position and sharpness of the maximum of the distribution (smaller $\alpha$ corresponding to broader maxima and closer to the origin).

Qualitatively we can say that, for the intensity, at low concentrations the distributions are of the Rayleigh type (with a maximum at x=0, and a long tail with an exponent $\mu$ very close to one, but for higher filling fractions a peak develops at non zero values of the intensity. As the coverage increases the position of this peak moves from zero to one, and $\mu$ reaches a value closer to 1.5. In all cases the transmission presents a much sharper peak than the intensity, in agreement with theoretical predictions[13]. As the coverage increases this peak becomes narrower, showing a much smaller range of dispersion of the values (compare the two transmissions illustrated in figure 2). This is reflected in the higher values of $\alpha$ attained, while $\mu$ remains in the neighborhood of unity. In all cases the conductance presents a behavior much more similar to the intensity than to the transmission, particularly in the case of randomized incoming phases. At low coverage it presents broad distributions peaking at the origin, which evolve into sharper peaks moving towards higher values as the coverage increases, accompanying the trend of the intensity distributions. The $\alpha$ values attained are closer to the intensity values, but $\mu$ remain consistently close to 1.5 (non-randomized phases) or to unity (randomized phases). In this last case the broad tails of the distribution are very different from the sharp distribution obtained for the transmission.

All the previous observations are based on the study of systems with filling fractions in the range 10%-60%. We also studied systems of high coverage (70%-90%). Because of the greater difficulty of obtaining the random samples of non-overlapping scatterers at such high filling fraction, we resorted to a different algorithm to generate them. The results obtained, while they cannot be quantitatively compared with the ones obtained in the range 10%-60%, are completely consistent with them and show the same growing trends with concentration.

**ANALYSIS OF THE MOMENTS**

The $n^{th}$ moment of a variable distributed according to $Y_U(x)$ is given by:

$$<x^n> = \frac{\Gamma^{n-1}(\frac{\alpha+1}{\mu})\Gamma(\frac{\alpha+n+1}{\mu})}{\Gamma^n(\frac{\alpha+2}{\mu})}$$

In the specific case of $\mu=1$ (all fits yielded $0.8<\mu<1.6$), this

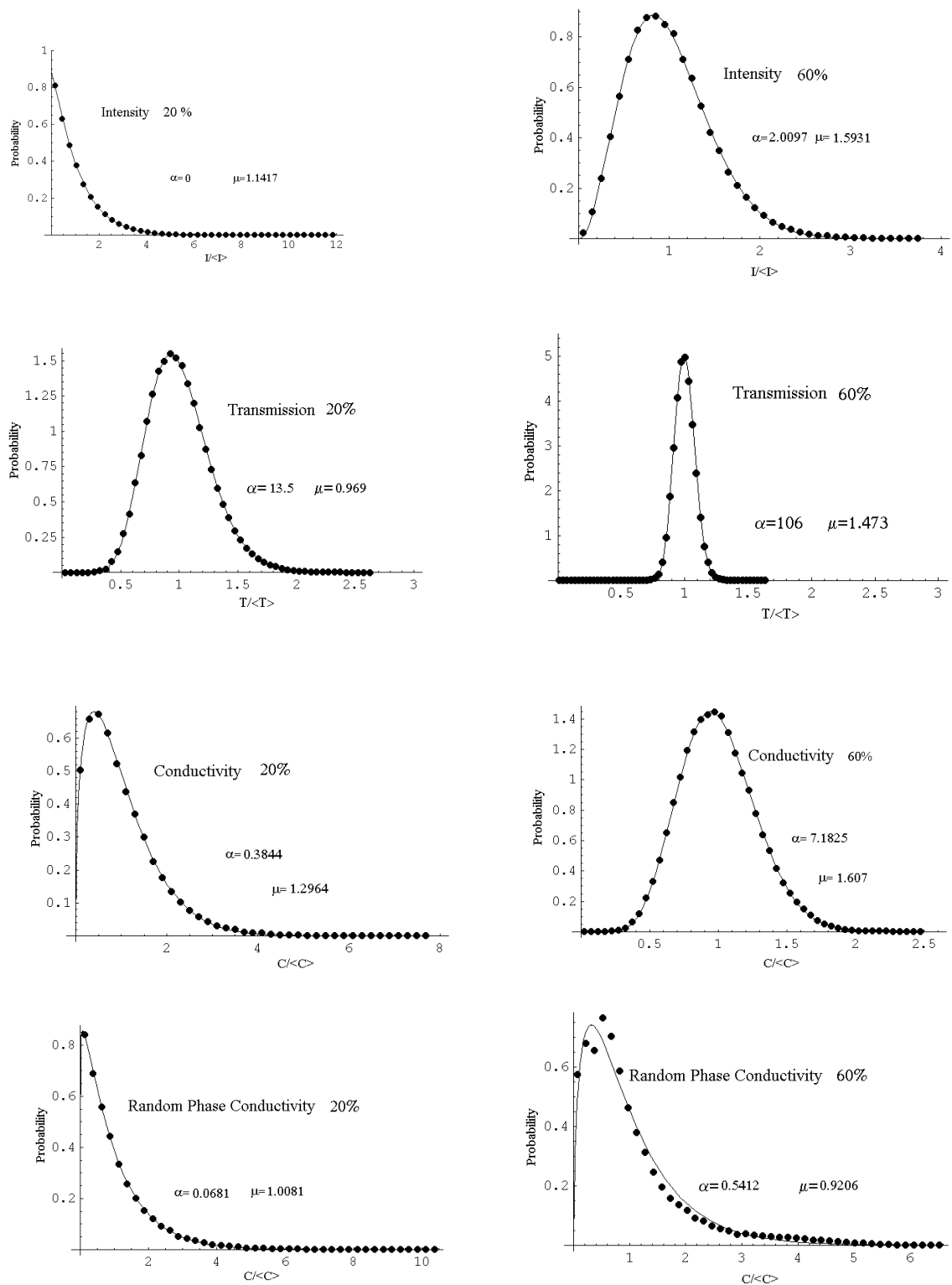

Figure 2. Probability distribution function for all transmittance quantities at low and high concentrations.

expression takes the particularly simple value:
$$\left(\frac{\alpha+n}{\alpha+1}\right)\left(\frac{\alpha+n-1}{\alpha+1}\right)\left(\frac{\alpha+n-2}{\alpha+1}\right)...\left(\frac{\alpha+2}{\alpha+1}\right)$$
which clearly shows that in the limit $\alpha \to 0$ we obtain n!, which is characteristic of the moments of Rayleigh distribution, while in the limit of large α these factors become progressively smaller tending to unity as in the case of the moments of a delta-function centered around one. Indeed, in our study the intensity distributions correspond always to small values of α, indicative of its long tails and fast growing moments, while the transmission distributions have bigger values of α, characteristic of narrower distributions, as expected. Another way of illustrating the same effect is to consider the rate of increase of consecutive moments,
$$\frac{<x^n>}{<x^{n-1}>} = \frac{\Gamma(\frac{\alpha+1}{\mu})\Gamma(\frac{\alpha+n+1}{\mu})}{\Gamma(\frac{\alpha+2}{\mu})\Gamma(\frac{\alpha+n}{\mu})}$$
which for the case μ=1 becomes $\frac{(\alpha+n)}{(\alpha+1)}$. This should be compared with the rate of increase in the case of Rayleigh distribution (α=0), which is simply *n*.

Several theoretical predictions[13] claim that the moments of the intensity and the transmission should be connected, in the diffusive regime by the relationship
$$\frac{<I^n>}{<T^n>} = n!$$
We have tested this hypothesis in our model. For each concentration we find that the ratio between the moments obtained from the distributions above is clearly smaller than n! As the concentration of scatterers increases beyond 20%, the discrepancy becomes more significant as seen in figure 3. There we show the logarithms of the ratios for the first ten moments according to the following expression

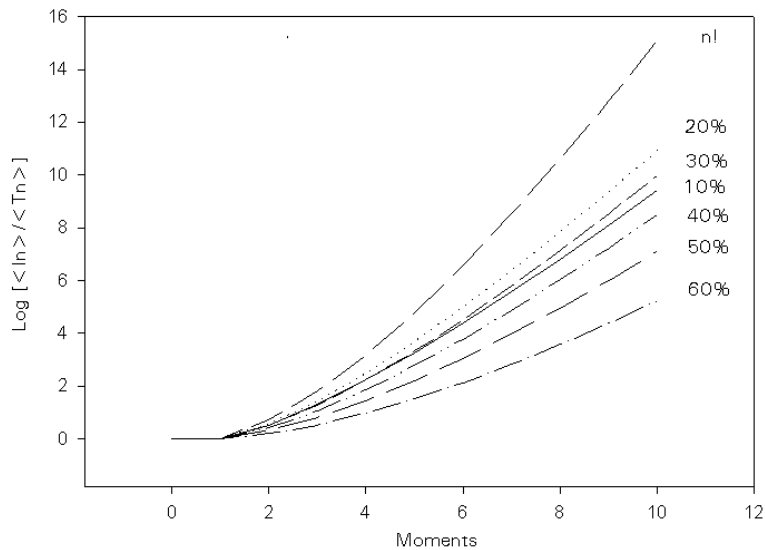

Figure 3. Ratios between moments of Intensity and Transmission

$$Log\left[\frac{<I^n>_{Y_U(\alpha_I,\mu_I)}}{<T^n>_{Y_U(\alpha_T,\mu_T)}}\right] = Log\left[\frac{\Gamma^{n-1}(\frac{\alpha_I+1}{\mu_I})\Gamma(\frac{\alpha_I+n+1}{\mu_I})}{\Gamma^n(\frac{\alpha_I+2}{\mu_I})}\right] - Log\left[\frac{\Gamma^{n-1}(\frac{\alpha_T+1}{\mu_T})\Gamma(\frac{\alpha_T+n+1}{\mu_T})}{\Gamma^n(\frac{\alpha_T+2}{\mu_T})}\right]$$

where $(\alpha,\mu)_{I(T)}$ stands for our best adjustment of the parameters $(\alpha,\mu)$ for the case of intensity and transmission.

In all cases the moments of the Conductance grow faster than those of Transmission and even those of Intensity. This means that the distributions of Conductance recently considered for a one-dimensional disordered system[21] are relatively broad as seen in figure 2. This is connected to the random interference between the incoming channels. This is particularly noticeable in the case of the random phase Conductance. There the moments of the various distributions remain close as the concentration increases ($\alpha$ and $\mu$ do not change significantly).

## ACKNOWLEDGEMENTS

We thank Sioan Zohar for his assistance in developing the random sample generation algorithm and running some of the simulations, and A. Z. Genack for useful discussions. This work is supported by the National Cancer Institute grant CA77796-02 and Yeshiva University.